\documentclass[reprint,twocolumn,notitlepage,prb,superscriptaddress,showpacs]{revtex4-1}
\usepackage{hyperref}
\usepackage[usenames,dvipsnames]{color}
\usepackage{amsmath}
\usepackage[english]{babel}
\usepackage{amssymb}
\usepackage{graphicx}
\usepackage{epstopdf}
\usepackage{comment}
\usepackage{mathrsfs}
\usepackage{amsfonts}
\usepackage{tabularx}
\usepackage{multirow}
\usepackage[]{todonotes}
\usepackage[]{algorithm2e}

\begin{document}
\title{Optimal grouping of arbitrary diagrammatic expansions via analytic pole structure}

\author{Amir Taheridehkordi}
\affiliation{Department of Physics and Physical Oceanography, Memorial University of Newfoundland, St. John's, Newfoundland \& Labrador A1B 3X7, Canada} 
\author{S. H. Curnoe}
\affiliation{Department of Physics and Physical Oceanography, Memorial University of Newfoundland, St. John's, Newfoundland \& Labrador A1B 3X7, Canada} 
\author{J. P. F. LeBlanc}
\email{jleblanc@mun.ca}
\affiliation{Department of Physics and Physical Oceanography, Memorial University of Newfoundland, St. John's, Newfoundland \& Labrador A1B 3X7, Canada} 

\date{\today}

\begin{abstract}
We present a general method to optimize the evaluation of Feynman
diagrammatic expansions, which requires the automated symbolic assignment
of momentum/energy conserving variables to each diagram. With
this symbolic representation, we utilize the pole structure of each diagram
to automatically sort the Feynman diagrams into groups that are likely to
contain nearly equal or nearly cancelling diagrams, and we show that
for some systems this cancellation is \emph{exact}. This allows for a
potentially massive cancellation {\em during} the numerical integration of
internal momenta variables, leading to an optimal suppression of the `sign
problem' and hence reducing the computational cost. Although we define
these groups using a frequency space representation, the equality or
cancellation of diagrams within the group remains valid in other
representations such as imaginary time used in standard diagrammatic Monte
Carlo. As an application of the approach we apply this method, combined
with algorithmic Matsubara integration (AMI) [Phys. Rev. B 99, 035120 (2019)] and Monte Carlo methods, to the Hubbard model self-energy expansion on a 2D square lattice up to sixth order
which we evaluate and compare with existing benchmarks.
\end{abstract}

\maketitle

\section{Introduction} 
One of the most challenging problems in condensed matter physics is correctly evaluating electronic interactions for free particle or lattice systems with many electrons. This problem is of course a subset of a more general problem, that of fermionic particles interacting through bosonic exchange. 
In one sense, this problem is addressed by many-body perturbation theory
using the formalism of Feynman diagrammatics, which allows one to construct in an
intuitive manner the contributions at each order in perturbation theory.\cite{Feynman,Baym,Luttinger:1960}
In practice, however, it is extraordinarily difficult to handle more than
just the lowest order diagrams due to the factorial increase\cite{Kugler} in the number of
diagrams at each order, and this is further exacerbated by the high dimensional integrals that must be performed in order to evaluate each Feynman diagram.

Diagrammatic Monte Carlo (DiagMC) provides a powerful tool to numerically evaluate such diagrammatic expansions.\cite{Prokof'ev:1998,vanhoucke,vanhoucke:natphys,kozik:2010,rossi2017determinant}  However,  there is in general a Monte Carlo sign problem\cite{Loh,Chandrasekharan} with multiple origins. The first, warmly referred to as the `sign blessing',\cite{signblessing} is the huge cancellation that must exist between different Feynman diagrams at each order in order for the series to converge. The second sign problem occurs during the integration of each individual diagram, since the integrand in frequency space does not have a definite sign.
One can devise methods to mitigate the second problem, but the first, the
cancellation between topologically distinct diagrams, is disastrous to standard DiagMC.  Recently, a number of proposals to address this have surfaced, such as grouping diagrams based on some criterion with the hope of cancellation,\cite{Chen2019} or reconstructing the expansion in the form of a determinant.\cite{rossi2017determinant,simkovic2017determinant, alice}  These methods rely on the Matsubara formalism in that final results are evaluated for Matsubara (imaginary) frequencies ($i\nu_n$) or imaginary times ($\tau$) and not on the real frequency axis. This has excellent utility for thermodynamic properties where the temporal degree of freedom is integrated, but is problematic for direct frequency dependent observables such as the density of states since the analytic continuation from $i\nu_{n}\to\nu+i0^+$ cannot be uniquely performed for numerical data and requires the ill-posed inversion via methods such as maximum entropy inversion.\cite{Levy2016, jarrell:maxent} Performing such procedures ultimately dominates the uncertainty in the result and undermines any attempt at precision numerics.\cite{Mravlje}  Worse still is the compression of Matsubara frequencies in the $T\to0$ limit where numerical Monte Carlo methods become effectively non-ergodic leading to poor convergence.

The entirety of this problem can be sidestepped by simply following standard many-body theory and evaluating the internal Matsubara sums analytically.  The only roadblock to doing so is the complexity of the analytic equations.  This roadblock has recently been overcome by the method of algorithmic Matsubara integration (AMI)\cite{AMI} that in principle allows for the symbolic evaluation of the Matsubara sums for arbitrarily complex Feynman diagrams with minimal computational expense. The analytic result of AMI can be evaluated at \emph{any} temperature and the analytic continuation is trivialized since it can be imposed symbolically: $i\nu_{n} \to \nu + i0^+$.
What remains is to sample a factorially growing space of diagram topologies and perform the spatial integrals, a problem typically reserved for DiagMC.  However, since AMI is formulated on the frequency axis, standard DiagMC will suffer a severe sign problem, and cannot be directly implemented.

In this manuscript we take a new approach to the sign problem. By considering Feynman diagrams in the Matsubara frequency representation we define a general procedure based on the analytic structure of each integrand that allows us to identify sets of topologically distinct Feynman graphs that \emph{exactly cancel} or are \emph{exactly equal} and further to identify other diagrams that can be trivially evaluated to be zero. In systems where the cancellation is not exact our method can identify and pair \emph{nearly} cancelling diagrams,
i.e., we systematically construct optimally sign-blessed groups. By pairing such diagrams during the numeric evaluation of momenta integrals we guarantee a huge cancellation, which suppresses the sign problem. 
Our procedure is general in that it can be applied to any Feynman diagrammatic expansion with any interaction.
As proof of principle we evaluate the numerical results for a particular  perturbative expansion, the Hubbard model \cite{Hubbard1963,benchmarks} on a two-dimensional square lattice. We construct the self-energy perturbative expansion up to sixth order at and away from half-filling. We then systematically group diagrams to provide what we believe to be the optimal set of diagrams to be evaluated using AMI and Monte Carlo methods and compare low order results to other numerical methods and benchmarks.

\section{Methods} \label{sec:methods}
In this section we outline each step required to group and evaluate diagrammatic expansions. What we propose is in fact conceptually simple but notationally complicated and for this reason we take a pedantic approach and describe in detail how to: generate and store diagrams symbolically; automate the evaluation of Matsubara sums analytically via AMI; how to systematically classify diagrams and construct the optimal groups of diagrams for a particular problem.

Central to doing this is the pre-generation of diagrams and assignment of symbolic momentum conserving variables, the first step in the standard procedure for translating Feynman graphs to integrals. This is not typically done in DiagMC, which instead probes energy/momentum configurations via the propagation of worms.\cite{worms} We will see that how momentum conserving variables are assigned is not unique, but that each diagram contains fundamental pole structure that cannot be hidden via some obscure choice of momentum conserving variables. We therefore base our diagram classification on these fundamental and physical poles.
We will typically describe the application to the self-energy expansion but in general the same procedures can easily be extended to other multileg or bosonic particle expansions.

\subsection{Constructing Diagrams and Integrands}
\begin{figure}
\centering
\includegraphics[width=0.8\linewidth]{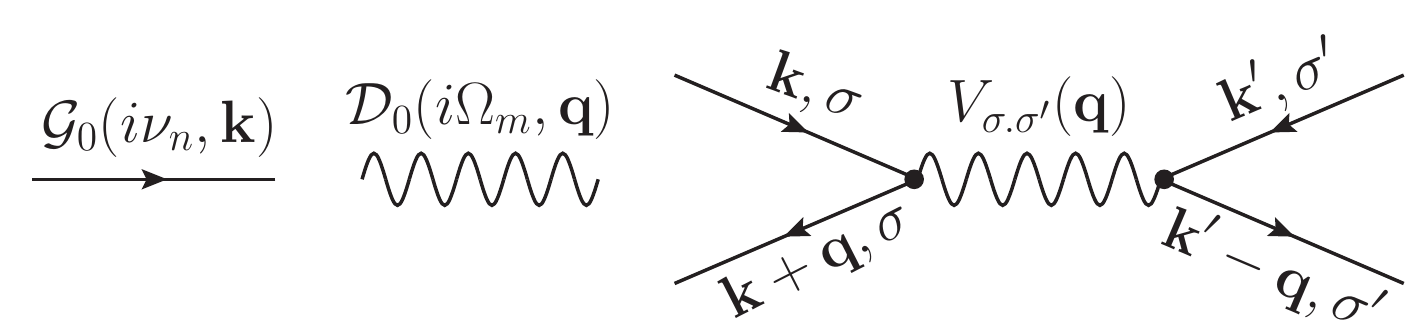}
\caption{Diagrammatic representation of \emph{Left}: Fermionic line, \emph{Middle}: Interaction line, and \emph{Right}: Two-body interaction between two fermionic lines with spins $\sigma$ and $\sigma'$ for two-body interaction $V_{\sigma,\sigma'}(\textbf q)$. Each line should be assigned with momentum/energy conserving variables. \label{fig:int_diag}}
\end{figure}
The building blocks of a Feynman diagram are fermionic (solid) and bosonic (wavy) lines (see Figure~\ref{fig:int_diag}). If there is a boson mediated two-body interaction $V_{\sigma,\sigma'}(\textbf q)$ in the system one can represent the interaction between two fermionic lines as shown in Figure \ref{fig:int_diag}, which additionally contains two factors of $\alpha$, the bare vertex. 
 Assuming one knows the free particle dispersion of each propagator and how they couple  (the bare vertex) then we have all the information required to convert the diagram into an integral. 
If one can first draw all possible topologically distinct diagrams up to a given interaction order (or number of loops) then the problem is essentially reduced to the evaluation of a set of integrals with known integrands. While stating this is simple, as already mentioned this is extremely challenging primarily due to the high dimensionality of the integrals.  

In order to systematically produce all terms in an expansion one requires a minimal set of procedures to change the order and topology of the diagrams. 
The simplest possible procedures are a set of two order-increasing processes:
to add an interaction line (AIL); and to add a tadpole (AT) (see Figure~\ref{fig:topol_procedures}). 
Without loss of generality, in what follows we will consider $V_{\sigma,\sigma'}(\textbf q)$ to be a Coulomb interaction but note that a general bosonic propagator $\mathcal D_0(i\Omega_m,\mathbf{q})$ can be similarly treated. Further, we restrict our discussion to the diagram space with 2-external legs, with the intent of constructing the set of self-energy diagrams, the set of one-particle irreducible diagrams - but the procedure remains unchanged for other diagrammatic series.
To generate the series we start with the lowest order diagrams of order $m$ and by systematically applying AIL and AT we generate all the possible diagrams of order $m+1$. 
Double counting of topologically equivalent graphs is not allowed and one therefore needs to discard duplicate graphs through explicit graph-isomorphism comparison. For this, the formal graph representation of diagrams is essential and the isomorphism checks can be aided by a tree decomposition of the graph. \cite{Dujmović}
We then store all of the topologically distinct, non-isomorphic diagrams. We then iterate the procedure at each order to generate all diagrams in the expansion up to an arbitrary order.

For each diagram of order $m$ with topology $\zeta_m$ we follow the Feynman rules to construct a corresponding mathematical expression.  These rules of course are well known.  We emphasize that our goal in this manuscript is not only to apply those rules, but in fact to automate the entirety of the process.  Therefore, we carefully express here those rules, to orient the reader: 
\begin{figure}
\centering
\includegraphics[width=0.6\linewidth]{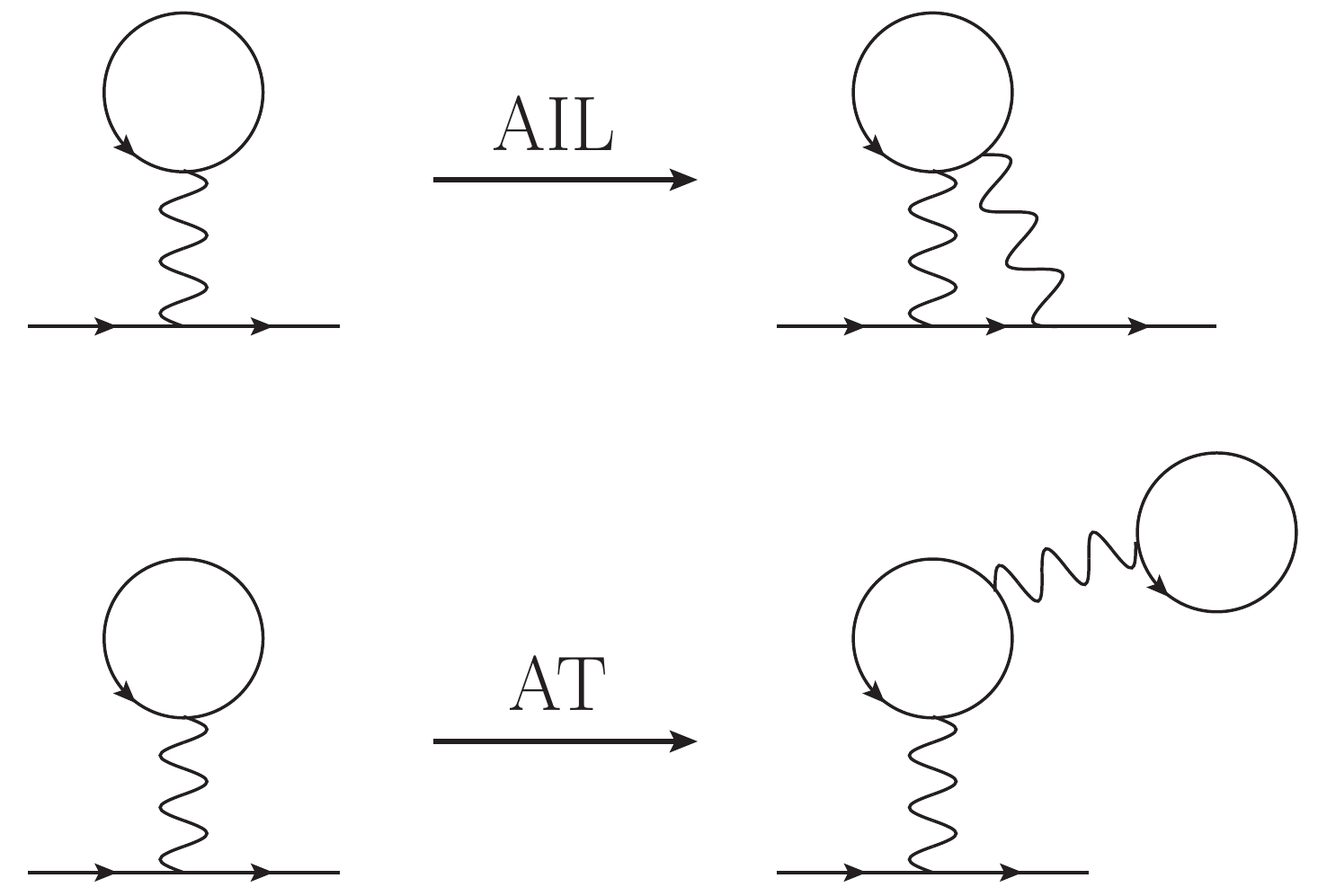}
\caption{Diagrammatic illustration of \emph{Top}: AIL and \emph{Bottom}: AT procedures in generation of the diagrammatic expansion described in the text. In the AIL one interaction (wavy) line is added to a current diagram while in the AT a tadpole (simple fermionic loop with a wavy tail) is added to the diagram. \label{fig:topol_procedures}}
\end{figure}
\begin{enumerate}
  \item We assign frequency-momenta  variables to each line such that conservation at each vertex is satisfied. We call a set of such variables for a given diagram a `label' and store it as an array. This procedure is outlined in Sec.~\ref{sec:labeling} and is not unique; it results in a number of distinct, but mathematically equivalent representations of the diagram integrand.  Non-unique labels are an issue that we will discuss in detail in Sec.~\ref{sec:labeling}.
  \item For a diagram with $N$ solid lines we assign bare Green's function $\mathcal{G}^j_0=\frac{1}{X^j-\epsilon^j}$ to each solid line $j$ with $j=1,2,...,N$, where $X^j$ represents the frequency, and $\epsilon^j$ the energy (see Sec.~\ref{section:formalism}).
    \item Each interaction line connecting two solid lines with spins $\sigma$ and $\sigma'$ should be directed and associated with $V_{\sigma,\sigma'}(\textbf q)$, where $\textbf q$ is determined via conservation rules at each vertex and is not an independent variable.
  \item Each internal Matsubara frequency and momentum is integrated.
  \item The result should be multiplied  by $\frac{(-1)^{m+F_{\zeta_m}}(2s+1)^{F_{\zeta_m}}}{(2\pi)^{nd}\beta^n}$ where $m$ is the order (number of the interaction lines) of the diagram, $F_{\zeta_m}$ is the number of fermionic loops in the diagram, $s$ is the spin, $d$ is the dimensionality of the system, $n$ is the number of independent frequencies, and $\beta$ is the inverse temperature.
  \item If the $j$th solid line closes on itself, i.e., a tadpole occurs, we insert a convergence factor, $e^{X^j 0^+}$.
\end{enumerate}

These rules are applied to each diagram resulting in (up to convergence factors) the Feynman integral for topology $\zeta_m$ at order $m$ in the form
\begin{align}
D_{\zeta_m}=\frac{1}{(2\pi)^{nd}\beta^n}\sum\limits_{\{\textbf k_n\}} \sum \limits _{\{\nu_n\}} &{A(m,s,F_{\zeta_m})} \prod
\limits_{j=1}^N \mathcal G_0^j(\epsilon ^j, X^j ) \times \nonumber \\ &\prod \limits_{m=1}^M V_{\sigma,\sigma^{'}}(\textbf q_m), \label{eqn:each_diag_in_goal}
\end{align}
where $N$ and $M$ are the number of fermionic (solid) and bosonic (wavy) lines in the diagram, respectively, $X^j$ is the frequency, $\epsilon ^j$ is the energy of the $j$th solid line, and $A(m,s,F_{\zeta_m})=(-1)^{m+F_{\zeta_m}}(2s+1)^{F_{\zeta_m}}$. Finally, an arbitrary diagrammatic expansion, $Q$, can be written down as the sum of each distinct diagram at each order
\begin{align}
Q(x_{ext})=\sum\limits_{m=0}^{\infty}\sum\limits_{\zeta_m} D_{\zeta_m}, \label{eqn:goal}
\end{align}
where the sum over $\zeta_m$ is over all unique topologies of order $m$. The result only depends on a set of external parameters, $x_{ext}$, which includes external frequency, external momenta, chemical potential and temperature of the system. 
\subsection{Evaluation of Matsubara Frequency Summations}\label{section:formalism}
Each diagram in the perturbative expansion defined by Eq.~(\ref {eqn:each_diag_in_goal}) consists of summations over Matsubara frequencies and over momenta within the first Brillouin zone. 
We perform the (unbounded) Matsubara sums by using algorithmic Matsubara integration (AMI) introduced in Ref.~\onlinecite{AMI}. The Matsubara summations of a given Feynman diagram $D_{\zeta_m}$ are contained in the factor
\begin{align}
I_{\zeta_m}=\frac{1}{\beta^n} \sum \limits _{\{\nu_n\}} \prod
\limits_{j=1}^N \mathcal G_0^j(\epsilon ^j, X^j )  \label{eqn:AMI_target}
\end{align}
of Eq.~(\ref{eqn:each_diag_in_goal}). Essentially, AMI is a procedure that evaluates the Matsubara sums by iteratively applying residue theorem to Eq.~(\ref{eqn:AMI_target}). We briefly review AMI here.

\subsubsection{Symbolic Array Representation of the Bare Green's Functions}
The first step in AMI is to represent the bare Green's functions $\mathcal G_0^j(\epsilon ^j, X^j )$ in a symbolic array form. We use the self-energy function as an example, which has one external frequency and one external momentum. 

For a given diagram $D_{\zeta_m}$ with $n$ independent (internal) Matsubara frequencies we define the frequency of each line as the linear combination $X^j=\sum_{\ell=1}^{n+1}i\alpha_\ell^j \nu_{\ell}$, where the allowed values for the coefficients $\alpha_\ell^j$ are zero, plus one or minus one. We store these coefficients as an array of length $n$, $\vec \alpha^j = (\alpha_1^{j}, ..., \alpha_{n}^{j})$ for the $j$-th solid line of the diagram. Similarly the free particle energy is $\epsilon^j = \epsilon(\textbf k_j)$, where $\textbf k_j=\sum_{\ell=1}^{n+1} \alpha_{\ell}^j \textbf k_{\ell} $. In this notation $\nu_{n+1}$ and  $k_{n+1}$  are the external frequency and momenta, respectively. 

We will also need to express $\epsilon^j$ in an array form. For a given diagram, the $j$th line (out of $N$ total solid lines) has an energy $\epsilon^j$, which will be one of $r$ symbolically different energies $e_\ell$ for $\ell \leq r$. This allows us to represent each distinct $\epsilon^j$ as an array with length $N$, where one entry takes the value $1$ and the rest are zero:
\begin{eqnarray}\label{eqn:dispersion_array_rep}
\epsilon^j = e_{\ell} \to \vec E_{\ell} = (\mathfrak {\delta}_{\ell,1}, \mathfrak {\delta}_{\ell,2}, ..., \mathfrak {\delta}_{\ell,N}),
\end{eqnarray}
where $\delta_{\ell,j}$ is a Kronecker delta. We are now able to represent each Green's function, $\mathcal G_0^j$, as an array with length $N+n$: \footnote{Note that the this array representation enables us to store both frequency and dispersions symbolically, while in Ref. \onlinecite{AMI} only frequency part is stored symbolically.}
\begin{eqnarray}\label{eqn:array_rep_single_G}
\mathcal G_0^j(\epsilon ^j, X^j) \to [\vec E_{\ell}, \vec \alpha^j].
\end{eqnarray}
This array representation (\ref{eqn:array_rep_single_G}) is equivalent to a full symbolic representation of each bare Green's function in the frequency-momenta space. 
\subsubsection{AMI Procedure}
\label{sec:AMI}
Eq.~(\ref{eqn:array_rep_single_G}) enables us to represent the product of the bare Green's functions in Eq.~(\ref{eqn:AMI_target}) as a nested array of size $N\times (N+n)$:
\begin{eqnarray}\label{eqn:Multi_Green_array_rep}
\prod_{j=1} ^ N \mathcal G_0^j(\epsilon^j, X^j) \to \bigg [[\vec E_{\ell_1}, \vec \alpha^1]; [\vec E_{\ell_2}, \vec \alpha^2]; ...; [\vec E_{\ell_N}, \vec \alpha^N] \bigg]. \nonumber \\
\end{eqnarray}
To clarify this, we provide in Supplemental Material the array representation of a particular third order self-energy diagram as an example. 

Starting with the array representation (\ref{eqn:Multi_Green_array_rep}) and following the AMI procedure\cite{AMI} we construct and store the AMI result. A typical AMI result contains many terms, which are represented as nested arrays. Each array contains two entries. The first entry is energy (momenta) part $E$, which is represented by an array $\vec {\mathcal E}$ which is a linear combination of $\vec E_{\ell}$ arrays defined by Eq.~(\ref{eqn:dispersion_array_rep}). Then the symbolic energy in general is constructed by $E=\sum_{\ell} \mathcal E_\ell e_{\ell}$, where the allowed values for $\mathcal E_{\ell}$, entries of $\vec {\mathcal E}$, are zero, plus one, or minus one. The second entry of the result is the frequency part, which is a linear combination of $\vec \alpha^j$ arrays. From the AMI result  the full symbolic result for Matsubara sums ($I_{\zeta_m}$ in Eq.~(\ref{eqn:AMI_target})) is obtained. Thus, for each diagram we have reduced the original problem of Eq.~(\ref{eqn:each_diag_in_goal}) to a sum over momenta:
\begin{align}
D_{\zeta_m}=\frac{1}{(2\pi)^{nd}}\sum\limits_{\{\textbf k_n\}} \mathfrak {D}_{\zeta_m}(i\nu_{n+1},\{\textbf k_{n+1}\},\beta,\mu), \label{eqn:goal_after_AMI}
\end{align}
where
\begin{align}
\mathfrak {D}_{\zeta_m}=A(m,s,F_{\zeta_m}) I_{\zeta_m}(i\nu_{n+1},\{\textbf k_{n+1}\},\beta,\mu)  \prod \limits_{m=1}^M V(\textbf q_m).\nonumber \\ \label{eqn:each_diag_after_AMI}
\end{align}
In summary, AMI enables us to analytically evaluate the Matsubara sums of all Feynman diagrams with minimal computational expense. 

\subsection{Classifying Diagrams via Pole Structure}
It has been pointed out that in order for the perturbative series to converge,\cite{signblessing} there must be a huge cancellation between topologically distinct diagrams at each order.  An unguided summation of diagram topologies therefore encounters a cataclysmic sign problem.\cite{kozik2015nonexistence,simkovic2017determinant,fedor:2019}
We seek to resolve this issue in this and the following sections where we introduce a systematic approach that allows us to identify these cancelling diagrams and remove them from the series. In addition, we find groups of equal diagrams, which provides us with a further reduction in computational cost. 

 Our goal is to evaluate Eq.~(\ref{eqn:goal}) truncated at a cutoff order $m_c$.  However, as we shall see, we do not really need in general to evaluate all the diagrams in the expansion; it turns out that some diagrams are individually vanishing. Furthermore, there exist diagrams that are exactly cancelling or equal. To this end we provide a filtering process to systematically identify the individually vanishing, as well as cancelling and equal diagrams, without any explicit evaluation of the frequency and momenta summations in Eq.~(\ref{eqn:each_diag_in_goal}). This allows us to substantially reduce the diagrammatic space of the problem leading to a significant reduction in the computational cost. In addition, since we eliminate the problematic vanishing and cancelling diagrams we markedly suppress the sign problem. 

There exist many expansions where \emph{nearly} cancelling diagrams appear, e.g., Hubbard self-energy diagrams away from half-filling. In order to manage the sign problem in these cases, we provide a general prescription in Sec.~\ref{sec:nearly_cancelling} to carefully treat the nearly cancelling pairs.
\subsubsection{Labeling Procedure}\label{sec:labeling}
As mentioned, the label of a Feynman graph is not unique and we need to carefully consider the role of labeling in this challenging problem. It is possible to generate the set of all labels for each diagram in the expansion from which one would construct the corresponding mathematical expressions using the Feynman rules. In the case of self-energy diagrams of order $m$ the number of independent (internal) Matsubara frequencies $n=m$, and the number of internal fermionic lines is $N=2m-1$. Thus, knowing the order of a diagram is sufficient to provide a complete accounting of possible labels.

In order to generate each label, we first assign the $n$ independent frequencies to a set of internal fermionic lines. We then assign dependent frequencies via conservation of energy at each vertex. If the conservation law at each vertex is satisfied then a valid label has been found. We generate all the possible labels by systematically choosing $n$ independent lines from $N$ possible choices.

Of course the number of energy conserving labels grows very fast with order, for example, while this number for the fourth order diagrams is of order 10, it is of order 100 by sixth order diagrams making this process more difficult with increasing order. 
Although expensive this labeling only needs to be performed once.  Further, having a symbolic representation of the labels
provides us with an opportunity to analytically extract the poles in the Matsubara frequency space, which as we shall show, plays a crucial role in identifying equal and cancelling diagrams. 
\subsubsection{Diagram Classes}
We are interested in identifying sets of diagrams that are either exactly equal or exactly cancel without performing the Matsubara and momenta sums. To begin, we propose to classify diagrams according to the pole structure of their integrands. One may recall that poles of the Green's functions have a physical manifestation as quasiparticles. If two diagrams are to be analytically equivalent (up to a sign), then they must contain the same set of non-removable divergences (virtual quasiparticles) in  order to produce the same integral result. This can only be true if the original integrands have the same pole structure. 

We define the \emph{pole configuration} of a diagram  $D_{\zeta_m}$ to be a set of integers $(n_1,n_2,\cdot \cdot \cdot,n_{max}$), where $n_i$ is the 
number of poles with multiplicity $i$ in the Matsubara frequency space and ${max}$ is the highest possible multiplicity of poles. Clearly, $\sum_{i=1}^{max}n_i=N$, the number of internal fermionic lines. It is important to note that the pole configuration does not depend on the choice of label of a diagram. In this way, we partition the set $S^m$ of diagrams of order $m$ into label-independent subsets $C_i^m$ of diagrams with the same pole configuration. We refer to these subsets as diagram classes. We illustrate this schematically in the top part of Figure~\ref{fig:class_sub}. 

Since it is not possible for diagrams that belong to different classes to be equivalent (or cancel) we need only look \emph {within} classes for equal or cancelling diagrams.
\subsubsection{Diagram Subclasses}
Now that we have grouped diagrams according to their pole configuration into classes $C_i^m$, we search for subclasses containing equal or cancelling diagrams. To this end, we now consider the distinct choices of a diagram's label, since how one chooses to label a given diagram might obscure its analytic equality or negation to another diagram in its class. Thus, we need a stronger condition in order to establish subclasses. We postulate then a \emph{necessary} condition in order for two diagrams $D_1$ and $D_2$ to be equal or cancelling:
\begin{itemize}
    \item For \emph{any} chosen label of $D_1$ there must exist a representation of $D_2$, where the integrands are equivalent or cancelling.
\end{itemize}
This simple postulate leads to the logical conclusion that the total number of labels of $D_1$ and $D_2$ must be equal, or else the two diagrams cannot be equal or cancelling in general.
\begin{figure}
\begin{center}
\includegraphics[width=0.45\textwidth]{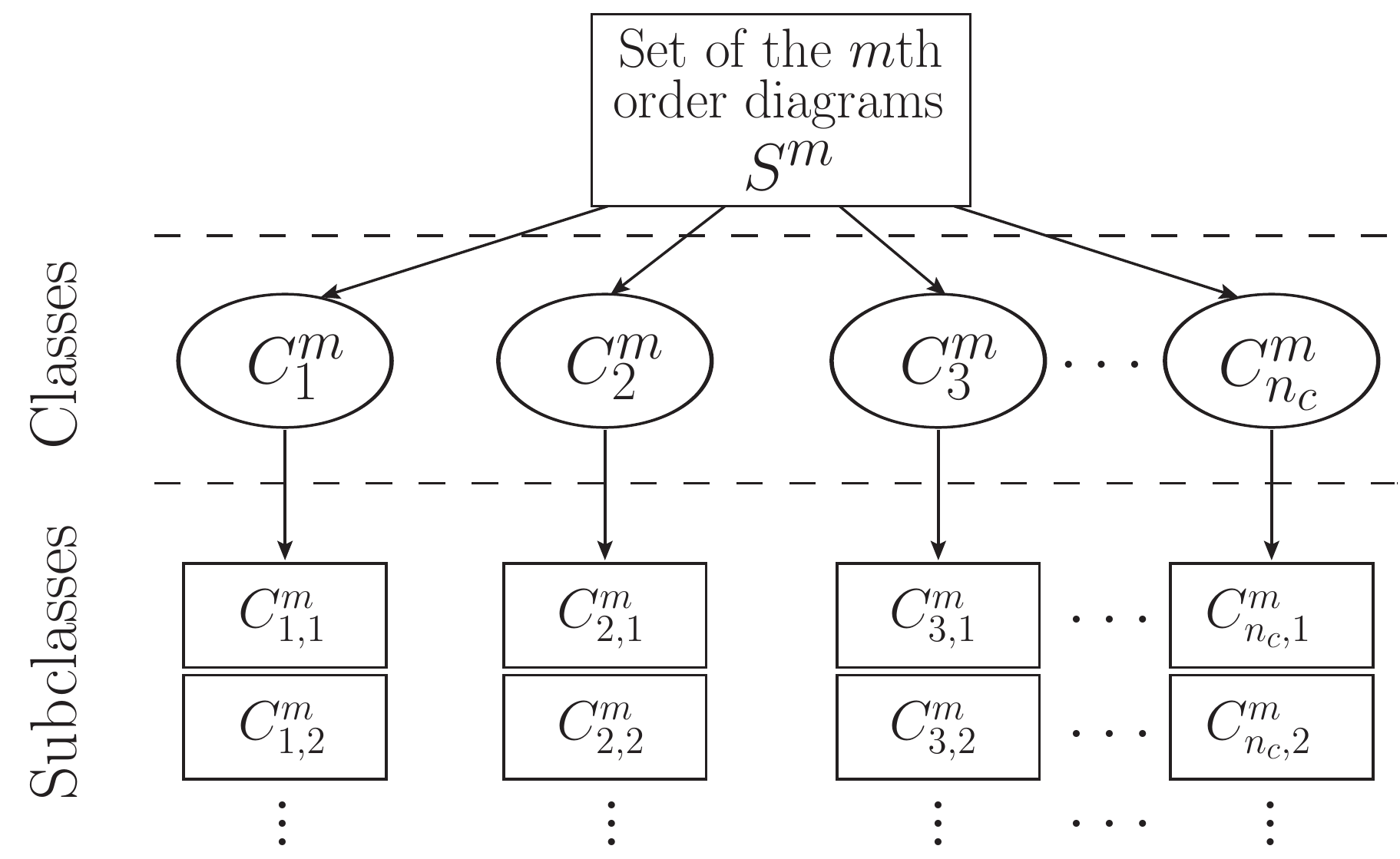}\\
\caption{Schematic illustration of classes and subclasses of the  set $S_m$ of diagrams of order $m$. Diagrams in a class $C_i^m$ have the same pole configuration, which are divided into subclasses of diagrams with similar characters. \label{fig:class_sub}}
\end{center}
\end{figure}
With this in mind 
we suggest a label-dependent identifier for a given integrand, which can be constructed by counting the number of poles with respect to each internal and external Matsubara frequency, $i\nu_i$, $x_i = \sum_{j=1}^N |\alpha_i^j|$. We then group these numbers into a set $(x_1,x_2,...,x_{n+1})$, which we then order the first $n$ entries from highest to lowest as $x=(x_i,x_j,x_k,...,x_{n+1})$ where $x_i\geq x_{j} \geq x_{k}$. (as above, we use self-energy diagrams, with one external frequency and $n$ internal frequencies, as an example). We call this object, $x$, the \emph{pole-ID} for a given integrand.
We now define \emph{diagram character} to be the complete set of pole-IDs generated by considering all possible labels of a diagram. Thus, the diagram character is label-independent.  We can therefore safely divide each class $C_i^m$ into subclasses $C_{i,j}^m$ of diagrams with the same diagram character. 
The bottom part of Figure~\ref{fig:class_sub} shows schematically the division of each class $C_i^m$ to subclasses $C_{i,j}^m$. 
\begin{figure}
\hspace*{-1.10cm} 
\includegraphics[width=0.50\textwidth]{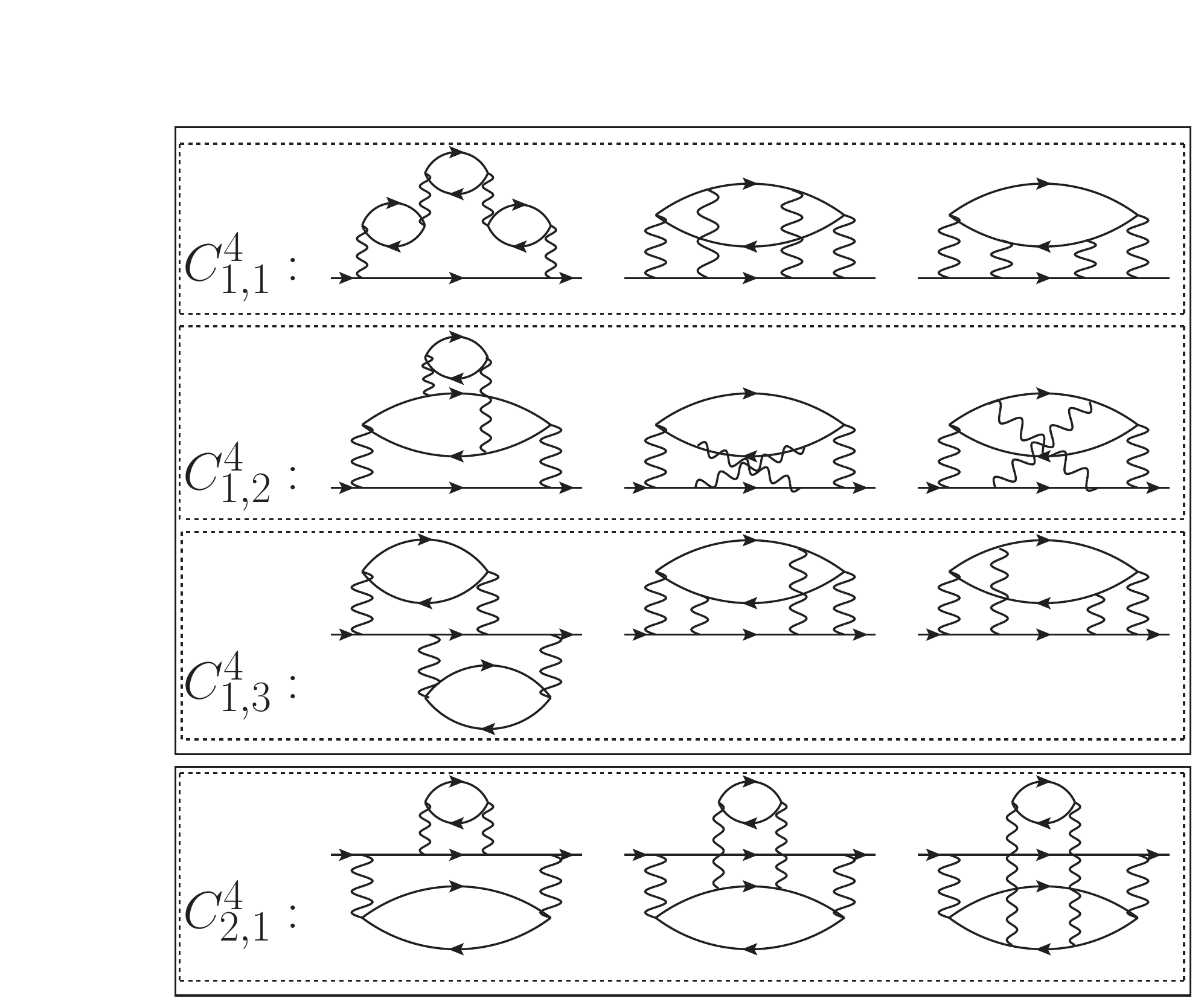}
\caption{Four subclasses of fourth order self-energy diagrams. The diagrams in the top panel belong to class $C_1^4$ with pole configuration (7,0), the diagrams in the bottom panel belong to class $C_2^4$ with pole configuration (5,1). Collecting pole-IDs for all possible labels, one finds out that the diagrams in each row have the same character, i.e., they belong to the same subclass. Thus, one takes the diagrams in each row as candidates to be either equal or cancelling.\label{fig:AID_4th}}
\end{figure}

As an example, we show in Figure~\ref{fig:AID_4th} twelve of the diagrams that contribute to the self-energy expansion at fourth order. These are divided into two classes, $C_1^4$ and $C_2^4$. Class $C_1^4$ further divided into three subclasses. One notes that in each row we observe a pair of diagrams that are isomorphic when one neglects the direction of the fermionic lines. We call such (non-isomorphic) diagrams, `almost isomorphic'. However, within each subclass there are also diagrams with wildly different topologies.
 
Since there can be no equal or cancelling diagrams that do not belong to the same subclass we need only compare diagrams belonging to the same subclass. Thus, the diagram character acts as a unique barcode or fingerprint that can be used to quickly group diagrams. When the number of graphs in a subclass is large, further filtering can be helpful, which we discuss next.

\subsection{Diagram Filter}\label{sec:filters}
Up to now we have only postulated that diagrams belonging to the same
subclass are likely to be equal or
cancelling. We can identify those diagrams which are equal up to a sign by applying
transformations to
their integration variables such that their integrands remain equivalent.
To this end, we introduce a filter, which we call {\em graph
invariant transformations}
(GIT). GIT identifies vanishing, cancelling and equal diagrams
within the subclasses.

\subsubsection{Graph Invariant Transformations (GIT)}
We begin the GIT procedure by  selecting a pair of diagrams $D_1$ and $D_2$. We then choose {\em one} label for each of $D_1$ and $D_2$ with
their integrands
stored in the array
representations (described in Sec.~\ref{sec:AMI}), which we call $L_1$ and $L_2$.
Next we will apply transformations to one of the labeled integrands and look for equality/negation.  These transformations must be such that they change the integrand but not the integral over internal parameters.

We identify three important transformation types.
The set of transformations $\mathcal T_1$
swaps two of the independent Matsubara frequencies,
\begin{eqnarray}\label{eqn:T_1}
\mathcal T_1: (i\nu_p,\textbf k_p) \leftrightarrow (i\nu_{p'},\textbf k_{p'}).  
\end{eqnarray}
We note that $\mathcal T_1$ 
is equivalent to a relabeling of the diagram that guarantees a new momentum conserving label.
The second transformation  $\mathcal T_2$ flips the sign of one of the internal
fermionic frequencies
and corresponding momentum,
\begin{eqnarray}\label{eqn:T_2}
\mathcal T_2: (i\nu_p,\textbf k_p) \to (-i\nu_p,-\textbf k_p).  
\end{eqnarray}
Finally, for many problems there might be another transformation $\mathcal T_3$ under which the dispersion of (at least) one of the solid lines changes sign:
\begin{eqnarray}\label{eqn:T_3}
\mathcal T_3: \epsilon^j \to -\epsilon^j. 
\end{eqnarray}

We apply the group of all possible transformations (including combinations
of $\mathcal T_1$, $\mathcal T_2$ and $\mathcal T_3$) to the integrand of the diagram, with each result
stored as array representations. Diagram $D_1$ equals or cancels diagram
$D_2$
if there is a transformation $\mathcal T$ such that
$\mathcal T:L_1 = \eta L_2$ with $\eta=\pm1$.  In practice, our procedure compares $\mathcal T:L_1$ with $L_2$ after each transformation and stops when such a
transformation is found.

GIT also enables us to identify the vanishing diagrams. To do so we start by selecting a diagram $D$ with its integrand represented by the array $L$. Then we apply GIT to look for a transformation $\mathcal T$ such that $\mathcal T:L = - L$. If such a transformation is found the diagram $D$ is trivially vanishing. 

\subsubsection{Application of GIT}
\begin{figure}
\centering
\includegraphics[width=0.3\linewidth]{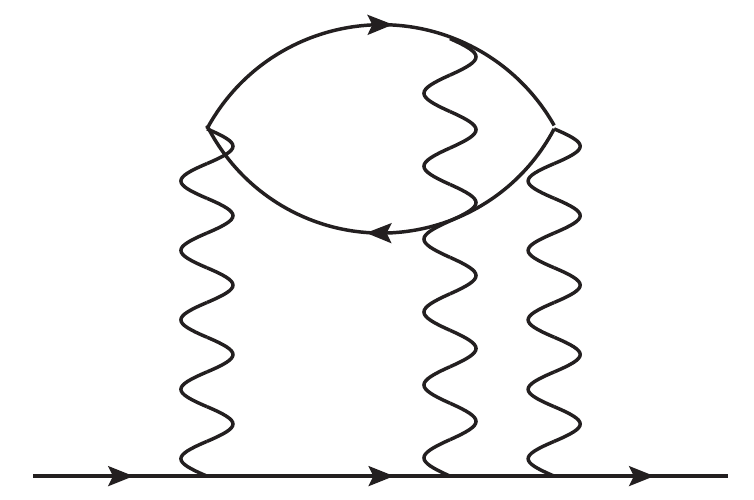}
\includegraphics[width=0.3\linewidth]{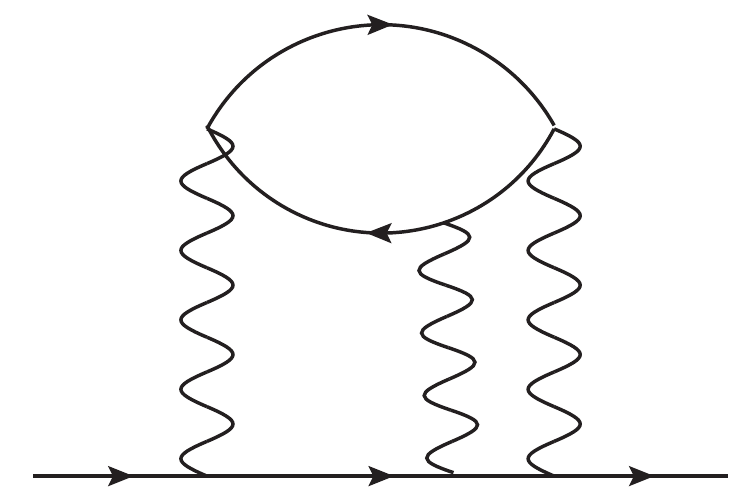}
\caption{Two topologically distinct third order self-energy diagrams, which are almost isomorphic. Application of GIT reveals that they are precisely cancelling at half-filling.   \label{fig:AI_example}}
\end{figure}

In order to identify vanishing, cancelling, and equal diagrams one needs to probe all the possible distinct pairs within each subclass by GIT. However, some considerations can substantially reduce the number of pairs to be investigated. First, for systems with particle-hole symmetry one can show that almost isomorphic diagrams are always either cancelling or equal. Therefore, one should first apply GIT to almost isomorphic diagrams since they are likely equal or cancelling. We show an example in Figure \ref{fig:AI_example} for two third order self-energy diagrams that are almost isomorphic. It is well known that for particle-hole symmetric systems these will cancel, and indeed application of GIT will find an appropriate set of transformations and return a value of $\eta=-1$. However, the same application to the subclasses of fourth order diagrams in Figure~\ref{fig:AID_4th} at half-filling will show that the diagrams are equal, and return $\eta=1$. Second, because comparison \emph{is} label dependent it follows that if the pole-IDs of two labeled graphs are not equivalent then the GIT can not show the equality or negation of the graphs for these specific labels. Thus, it is sufficient to only apply GIT to pairs of labeled diagrams with equivalent pole-IDs.

In practice, to apply the GIT procedure we first fix each diagram's label such that the number of times the external frequency appears in the label is maximized. This will allow us to find most but not all of the equal and cancelling diagrams. We then compare two diagrams and cycle through possible labels until we find a match (equal/cancelling) or exhaust all possible labels - they are not equal/cancelling. 

\subsection{Evaluation of Momenta Summations}
The final step in evaluating the diagrams in a perturbative expansion is
to perform the momenta integrations. Since these are high-dimensional
integrals one typically
uses  Monte Carlo integration. This will be efficient, if the number of diagrams in the expansion is not too large. In this approach we sample internal momenta points uniformly from $[-\pi,\pi]^{d}$ following standard Monte Carlo procedure. \cite{Zhiqiang} We generate $y$ samples in the internal momenta space each denoted by $\{\textbf p_n\}_i$ and approximate each diagram $D_{\zeta_m}$ by
\begin{align} \label{integral_naive_mc_gen}
D_{\zeta_m} \simeq D_{\zeta_m}^{(y)} = & \frac{(2\pi)^{nd}}{y} \sum_{i=1}^{y} \mathfrak {D}_{\zeta_m}(i\nu_{n+1},\{\textbf p_n\}_i, \textbf k_{n+1},\mu),  
\end{align}
from which the series expansion of $Q$ is calculated:
\begin{align}
Q\simeq \sum\limits_{m=0}^{\infty}\sum\limits_{\{\zeta_m\}} D_{\zeta_m}^{(y)}(i\nu_{n+1},\textbf k_{n+1},\beta.\mu). \label{eqn:goal_AMI_MNC}
\end{align}
For problems with a large diagrammatic space the direct evaluation of all the diagrams may be impractical and in that case one combines AMI with Metropolis-Hasting Monte Carlo (MHMC)\cite{Hastings} as in standard DiagMC \cite{vanhoucke} to numerically evaluate the momenta sums as well as probing different topologies in the expansion (\ref{eqn:goal}). This approach is similar to standard DiagMC but with three significant differences: First, we work in the Matsubara frequency space, similar to recent works on diagrammatic dual Fermion method, \cite{iskakov:2016,Gukelberger:DF} rather than imaginary time space. Second, we generate all the diagrams and their corresponding mathematical expressions before the MC simulation instead of producing the diagrams during the simulation. This, as we shall see, trivializes the detailed balance equations of a MHMC simulation. Third, since we analytically compute and store the Matsubara sums before MC simulation we eliminate the need for probing internal Matsubara frequencies (or equivalently internal imaginary times).        

To stochastically sample the diagrams we introduce a set of ergodic update procedures to probe diagram orders, diagram topologies, and internal momenta fixing all other external variables. For each step of the Monte Carlo simulation one of the updates is randomly chosen and the proposed configuration is accepted or rejected according to the Metropolis-Hastings scheme. We note that each diagram is identified by two elementary properties: order ($m$) and topology ($\zeta_m$). We assume that at order $m$ we have $\gamma_m$ different topologies in the expansion~(\ref {eqn:goal}). Now we introduce the following updates:
\begin{enumerate}
    \item Change momenta: The current momenta $\{\textbf k_n\}_c$ of the current diagram of order $m$ is changed to a proposed momenta $\{\textbf k_n\}_p$ where the $\{\textbf k_n\}_p$ are derived from the uniform distribution function $W(m)$.
    \item Change topology: By this update a diagram topology is changed within a specific order, i.e., if the current diagram is of order $m$ with topology $\zeta_m^c$ another diagram of order $m$ with topology $\zeta_m^p$ from the stored diagrams is proposed.
    \item Change order: The current diagram of order $m_c$ and with topology $\zeta_{m_c}^c$ is changed to a diagram of order $m_{p}$ and with topology $\zeta_{m_p}^{p}$. 
\end{enumerate}
Note that the proposed $m$th order topology is chosen uniformly from $\gamma_m$ possible topologies with probability $\frac{1}{\gamma_m}$. Finally, the acceptance probability of these updates in the Metropolis-Hastings scheme is expressed as
\begin{eqnarray}\label{detailed_balance_equation}
A=\hbox {Min} \bigg [1, \frac{\gamma_{m_p}}{\gamma_{m_c}} \frac{|\mathfrak {D}_{\zeta_{m_p}^p}^{re/im}(i\nu_{n+1},\{\textbf k_{n+1}\}_p,\beta,\mu)|W(m_c)}{|\mathfrak {D}_{\zeta_{m_c}^c}^{re/im}(i\nu_{n+1},\{\textbf k_{n+1}\}_c,\beta,\mu)|W(m_p)} \bigg], \nonumber \\
\end{eqnarray}
where $|\mathfrak {D}_{\zeta_m}^{re/im}|$ is the absolute value of the real/imaginary part of the $\mathfrak {D}_{\zeta_m}$ and $W(m)=1/(2\pi)^{md}$.

It is typical in MHMC to seek an update criterion that minimizes computational expense. Unfortunately, here one has no option but to evaluate the entire AMI integrand $\mathfrak {D}_{\zeta_m}$, which becomes expensive at high orders making it difficult to generate sufficient statistics.

\section{Example: Self-Energy for the 2D Square Lattice Hubbard Model} \label{sec:Hubbard}
As an application of our method, we calculate the self-energy for the
Hubbard model on a
two-dimensional square lattice up to sixth order in perturbation theory.
We consider the nearest neighbor tight binding dispersion given by $\epsilon(\textbf k) = -2t\big(\cos k_x + \cos k_y\big) -\mu$ where $t$ is the hopping amplitude and $\mu$ is the chemical potential. In this model, the potential is the momentum-independent local Hubbard interaction, $V_{\sigma,\sigma^{'}}(\textbf q)=U\delta_{\sigma,-\sigma^{'}}$.
The self-energy is
\begin{align}
\Sigma = & \sum\limits_{m=1}^{m_c}\sum\limits_{\{\zeta_m\}}\frac
{(-1)^{m+F_{\zeta_m}}U^m}{(2\pi)^{2m}\beta^m}\sum\limits_{\{\textbf k_m\}} \sum \limits _{\{\nu_m\}}& \prod
\limits_{j=1}^{2m-1} \mathcal G_0^j(\epsilon ^j, X^j ) \nonumber \\ & + O(U^{m_c+1}), \label{eqn:self_Hubbard}
\end{align}
evaluated to a cutoff order, $m_c$.
Here we remind the reader that a self-energy diagram of order $m$ has
$2m-1$ internal fermion lines
and $m$ independent frequencies and momenta.

Since the Hubbard interaction only occurs between fermionic lines with
opposite spins, we
construct and store only those connected one-particle irreducible diagrams
that satisfy this criterion. The total number of diagrams at each order $N_{init}^{(m)}$ is
given in the first row of Table \ref{tab:RF}. We then find all the possible labels (as explained in Sec.~\ref{sec:labeling}) for each stored diagram, which enables us to construct the classes and subclasses of self-energy diagrams.

\subsection{Diagrammatic Space Reduction for the Hubbard Self-Energy}
We first note that the contribution of diagrams with tadpole insertions (one-legged diagrams) can be neglected because they are equivalent to shifting the chemical potential $\mu \to \mu-\bar n U/2$, where $\bar n$ is the number of electrons per site. \cite{Zlatic,Daul1997} In doing so we in fact redefine chemical potential and self-energy function such that $\mu=0$ corresponds to half-filling. \cite{Gukelberger:2015}
As shown in Table \ref{tab:RF} this standard procedure substantially reduces the number of diagrams from $N_{init}^{(m)}$ to $N^{(m)}$. We then find all possible labels (as explained in  Sec.~\ref{sec:labeling}) for each diagram in
order to classify
the diagrams into subclasses; then we apply the GIT procedure within each
subclass to
identify vanishing, equal, and cancelling diagrams at half-filling.
\begin{table}
\caption{Diagrammatic space reduction by shifting the chemical potential for the Hubbard self-energy expansion up to sixth order. $N_{init}^{(m)}$: Total number of the $m$th order Hubbard self-energy diagrams in the original expansion. 
$N^{(m)}$: Total number of the $m$th order Hubbard self-energy diagrams neglecting all one-legged diagrams applying chemical potential shift.}
\label{tab:RF}        
\begin{tabular}{|c|c|c|c|c|c|c|}
\hline
$m$ & 1 & 2 & 3 & 4 & 5 & 6 \\
\hline \hline 
$N_{init}^{(m)}$ & 1 & 2 & 8 & 44 & 296 & 2312\\
\hline
$N^{(m)}$ & 0 & 1 & 2 & 12 & 70 & 515\\
\hline
\end{tabular}\\
\end{table}
\normalsize 

The transformations $\mathcal T_1$ and $\mathcal T_2$ are given by (\ref{eqn:T_1}) and (\ref{eqn:T_2}), respectively, and
the transformation $\mathcal T_3$ is a $(\pi, \pi)$ shift of internal momentum,
\begin{eqnarray}\label{eqn:T_2_Hubbard}
\mathcal T_3: \textbf k_p \to \textbf k_p+(\pi,\pi),  
\end{eqnarray}
which flips the sign of $\epsilon^j$ at half-filling if it depends on $\textbf k_p$. 
Since potential $U$ is a constant and the momenta sums are performed over the first Brillouin zone the expansion (\ref{eqn:self_Hubbard}) is invariant under any arbitrary combination of these transformations.
\begin{figure}
\begin{center}
\includegraphics[width=0.45\textwidth]{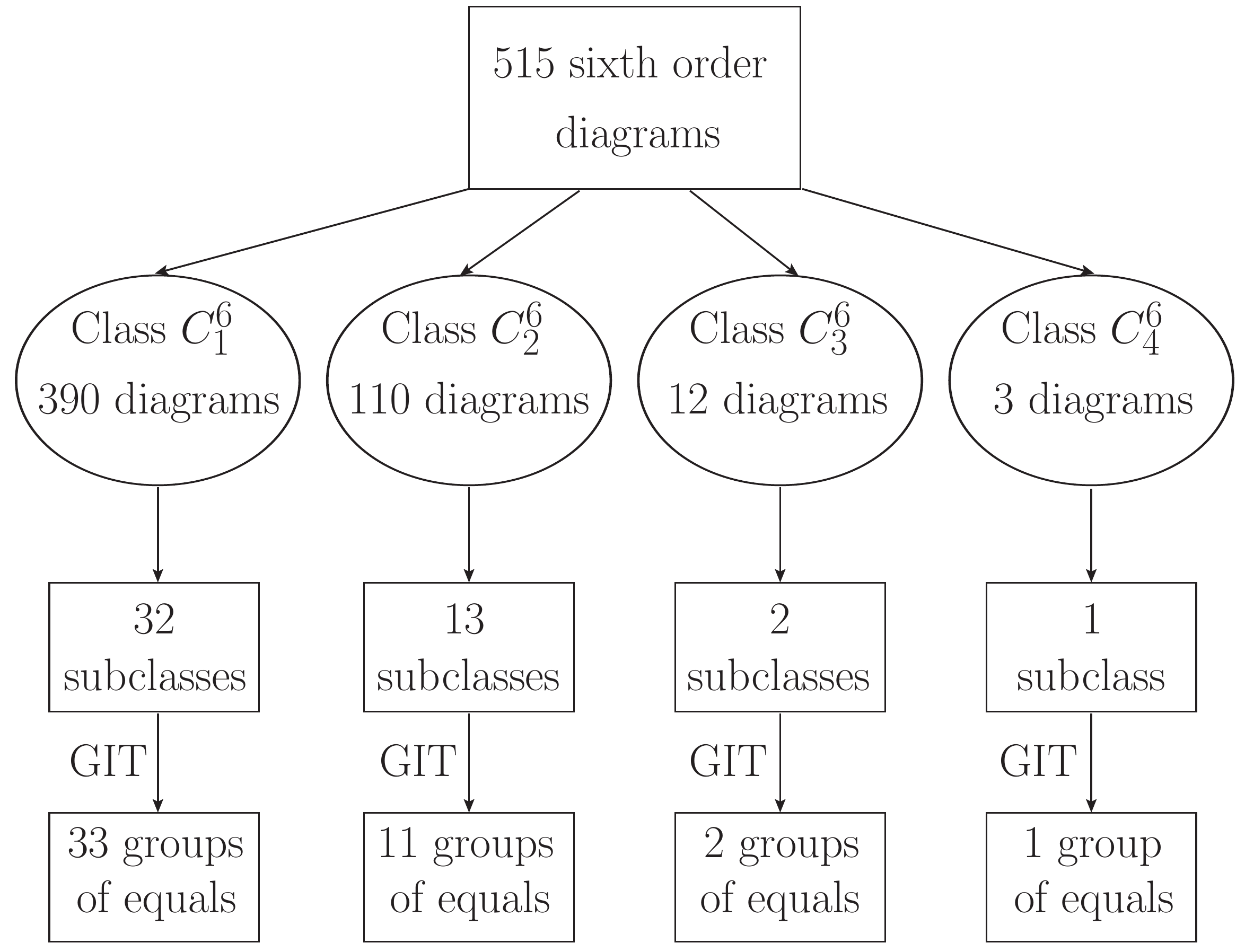}\\
\caption{Schematic illustration of constructing groups of sixth order equal diagrams at half-filling. We deal with 515 diagrams after chemical potential shift at sixth order. We first categorize the diagrams into classes according to their pole configurations. Note that $C^6_1$, $C^6_2$, $C^6_3$, and $C^6_4$ are classes of diagrams with pole configurations $(11,0,0)$, $(9,1,0)$, $(7,2,0)$, and $(8,0,1)$, respectively. We then construct the corresponding subclasses for each class considering their diagram characters. Finally, application of GIT within each subclass enables us to discard all the cancelling terms and find groups of equal diagrams at half-filling. \label{fig:6_order_groups}}
\end{center}
\end{figure}

In the third order there are only two diagrams, shown in Figure~\ref{fig:AI_example}. These two diagrams belong to the same subclass and the GIT procedure finds that they cancel.
In the Supplemental Material, we explicitly present the transformations
which relate these two diagrams. The twelve fourth order diagrams are categorized in two classes, which subdivide into a total of four subclasses containing three diagrams each (see Figure \ref {fig:AID_4th}). The GIT procedure reveals that the diagrams within each subclass are precisely equal in agreement with what has been reported previously. \cite{Freericks,Freericks2,Gebhard2003} There are 70 fifth order diagrams divided into 11 subclasses; GIT reveals that all of the diagrams within each subclass exactly cancel.
At sixth order we have 515 diagrams divided into four classes and
48 subclasses. Applying GIT, we identify 144 cancelling diagrams at half-filling; the remaining 371 diagrams are collected into 47 sets of equal diagrams. The details of the diagrammatic space
reduction for sixth order diagrams are illustrated in Figure~\ref{fig:6_order_groups}.  It is important to note that all precise cancellations found by the GIT procedure occur at half-filling only.

The diagrammatic space reduction for the Hubbard self-energy expansion
up to sixth order is summarized in Table~\ref{tab:summary}. There are no odd-order
diagrams in the reduced space, since each diagram
has a precisely cancelling partner; i.e., at half-filling odd order
diagrams do not contribute to the self-energy.\cite{werner} To calculate the
self-energy at half-filling, one needs to evaluate one diagram from each group, multiplied by the
number of diagrams in each group.  Altogether this represents a huge reduction: at sixth order
we began with 515 diagrams (not including one-legged diagrams); the GIT procedure
reduces this number to only 47 nonequivalent diagrams.
\begin{table}
\caption{Diagrammatic space reduction of the Hubbard self-energy up to
sixth
order at half-filling.  In the second row, $n_{tot}^{(m)}$ is the number of
subclasses at
each order $m$, and $(N^{(m)})$ is the total number of diagrams (not
included one-legged
diagrams) at each order $m$ (see Table~\ref{tab:RF}).
In the last row, $N_r^{(m)}$ is the number of groups of equal
diagrams at each order $m$, and $(n_d^{(m)})$ is the total number of non-cancelling
diagrams at each order $m$.}
\label{tab:summary}      %
\begin{center}
\begin{tabular}{|c|c|c|c|c|c|c|}
\hline
$m$ & 1 & 2 & 3 & 4 & 5 & 6 \\
\hline \hline
$n_{tot}^{(m)}(N^{(m)})$ & 0(0) & 1(1) & 1(1) & 4(12) & 11(70) & 48(515) \\
\hline
$N_{r}^{(m)}(n_d^{(m)})$ & 0(0) & 1(1) & 0(0) & 4(12) & 0(0) & 47(371)\\
\hline
\end{tabular}\\
\end{center}
\end{table}
\subsection{Sampling Nearly Cancelling Diagrams Away from Half-Filling}\label{sec:nearly_cancelling}
In practice {\em all} diagrams within each subclass should be stored in order to evaluate a given quantity away from half-filling. Diagrams which cancel at half-filling will nearly cancel away from half-filling, and the identification of those nearly cancelling
pairs can increase the efficiency of Monte Carlo integration.

The most straightforward way to evaluate diagrams away from half-filling is to sample the diagrams in each subclass as a whole instead of sampling diagrams one-by-one. However, to use the full power of the GIT in Monte Carlo integration away from half-filling one should group each nearly cancelling pair as a single integrand during the stochastic sampling. If a pair of diagrams $D_1$ and $D_2$ are exactly cancelling at half-filling, we essentially have a transformation $\mathcal T$ found by GIT and the necessary array representations $L_1$ and $L_2$, such that $\mathcal T: L_1 \to -L_2$ for every set of internal variables. One should then evaluate the pair of diagrams by considering $(\mathcal T:L_1)+L_2$ as a whole in the Monte Carlo sampling away from half filling. This optimizes the cancellation between the two diagrams. Thus, instead of sampling the nearly cancelling diagrams one-by-one, we sample them as a pair. This substantially improves the average sign and the uncertainty due to the huge cancellation.

\subsection{Numerical Results}
In this section we provide proof of concept results to illustrate the applicability of the method to the difficult problem of the Hubbard interaction. To do this we will first consider the order-by-order contributions for a point away from half-filling, both on the Matsubara and real frequency axis, in order to discuss the role of error induced by truncating the series. Subsequently we will compare our AMI calculations at half-filling to results from dynamical cluster approximation (DCA)\cite{leblanc:2013,benchmarks,fedor:2019, Maier05} as well as compare results from AMI on the real-frequency axis to those obtained via the numerical analytic continuation of DCA data. Finally we will compare the behavior on the Matsubara axis throughout the Brillouin zone to numerically exact results.

A central issue in truncated diagrammatic expansions for Hubbard interactions is that for large enough value of $U/t$ the truncated series is not convergent.\cite{rossi:2018}  To avoid this there are methods to improve convergence that in essence re-weight each diagram order without changing the sum of the entire series.\cite{wu:2017,simkovic2017determinant,Rossi:shiftedaction} Here, we would like to avoid any rescaling or resummation and instead we operate within the range of explicit convergence of the series.  
In order to do so, in each case we estimate our truncation errors (see Supplemental Material for derivation) by considering the behavior of coefficients at each order.  We can then use this information to evaluate the series for $U/t$ values such that the truncation error is small.
\begin{figure}
\centering
    \includegraphics[width=0.95\linewidth]{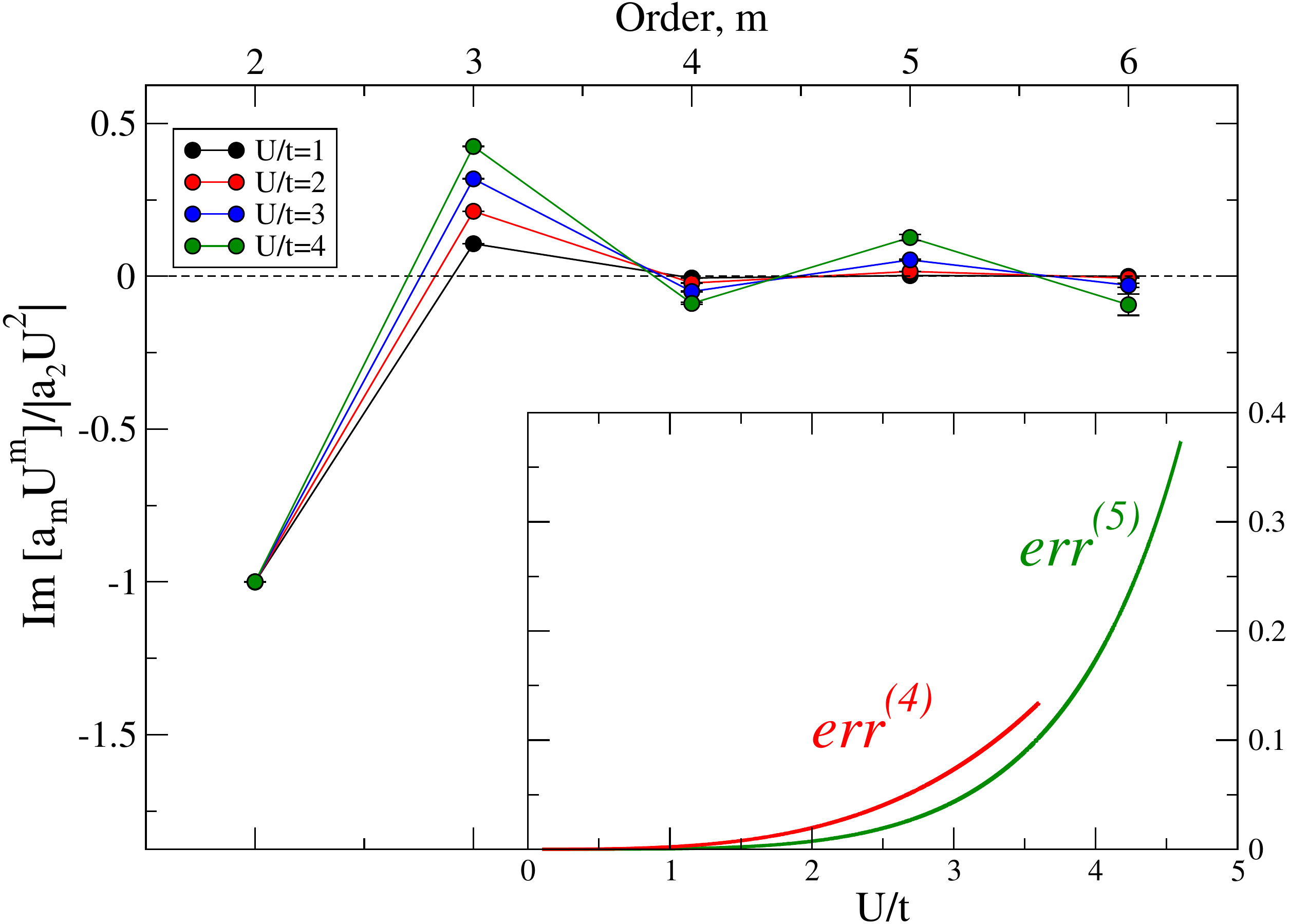}
        \caption{The contribution at each order to ${\rm Im}\Sigma_{\textbf k}(i\nu_0)$ for $U/t=1\to4$ normalized by the $m=2$ contribution.  Data are for parameters $\beta t=5$, $\mu/t=-1.5$ at $\textbf k=(\pi/8,\pi)$.\label{fig:doped}}
\end{figure}

As a first example,  shown in Figure~\ref{fig:doped},  we consider a case without particle-hole symmetry (i.e., away from half-filling), which means that all diagrams at each order (including odd orders) must be included.
Ignoring one-legged diagrams this represents the evaluations of $N^{(m)}$ diagrams at each order $m$ (the last row of Table~\ref{tab:RF}) where at each order the diagrams are grouped into $n_{tot}^{(m)}$ subclasses (see Table.~\ref{tab:summary}). 
The coefficients $a_m$ in the self-energy expansion $\Sigma = \sum_{m=2}^{6} a_m U^m + O(U^7)$ are evaluated for parameters $\beta t=5$, $\mu/t=-1.5$ at $\textbf k=(\pi/8,\pi)$ for $U/t=1$ to $4$. We normalize each plot by the absolute
value of the second order term. For $U/t \leq 2$ 
we see that the 4th, 5th and 6th order contributions are negligible;
however for $U/t=4$, these contributions are comparable to each other in
magnitude.
These findings are consistent with the truncation error, plotted in
the inset to Figure~\ref{fig:doped}.
The fractional truncation error $err^{(m)}$ is the error
estimate for truncating the series at order $m$, shown for $m=4,5$.
 We see that up to $U/t=2$ the truncation error is negligible ($err^{(5)}<1\%$) while at $U/t=4$ the truncation is $\approx 15\%$ and becomes divergent slightly above $U/t=4$.
The fourth order truncation error $err^{(4)}$  has only minor differences to
$err^{(5)}$.
This suggests that at this temperature ($\beta t = 5$) and values of
$U/t$ as large as 4, a diagrammatic series might be reasonably
approximated by neglecting terms higher than 5th order.
Actually, it is surprising that, for a wide
range of $U/t$ values, 4th or 5th order results should produce truncation errors $< 10\%$. Such behavior
has been observed at strong coupling from $\Sigma$DDMC,\cite{simkovic2017determinant,rossi2017determinant}
where the results of the diverging series at higher order
oscillate around the result such that the sum of all higher order terms is only a small contribution for weakly coupled cases though this ceases to be the case for large
values of $U/t$.

One should also note that to get reliable error bars for higher
order contributions, grouping the diagrams into subclasses is essential. In our example, at sixth order the average sign obtained by one-by-one
sampling of the diagrams in frequency space is $|\langle sign \rangle|<0.002$ making this problem intractable for standard DiagMC methods. However, by sampling within groups of diagrams, the average sign is improved to $\approx 0.11$. While still a very small average sign, this represents an enormous computational savings.  
\begin{figure}
\centering
    \includegraphics[width=0.95\linewidth]{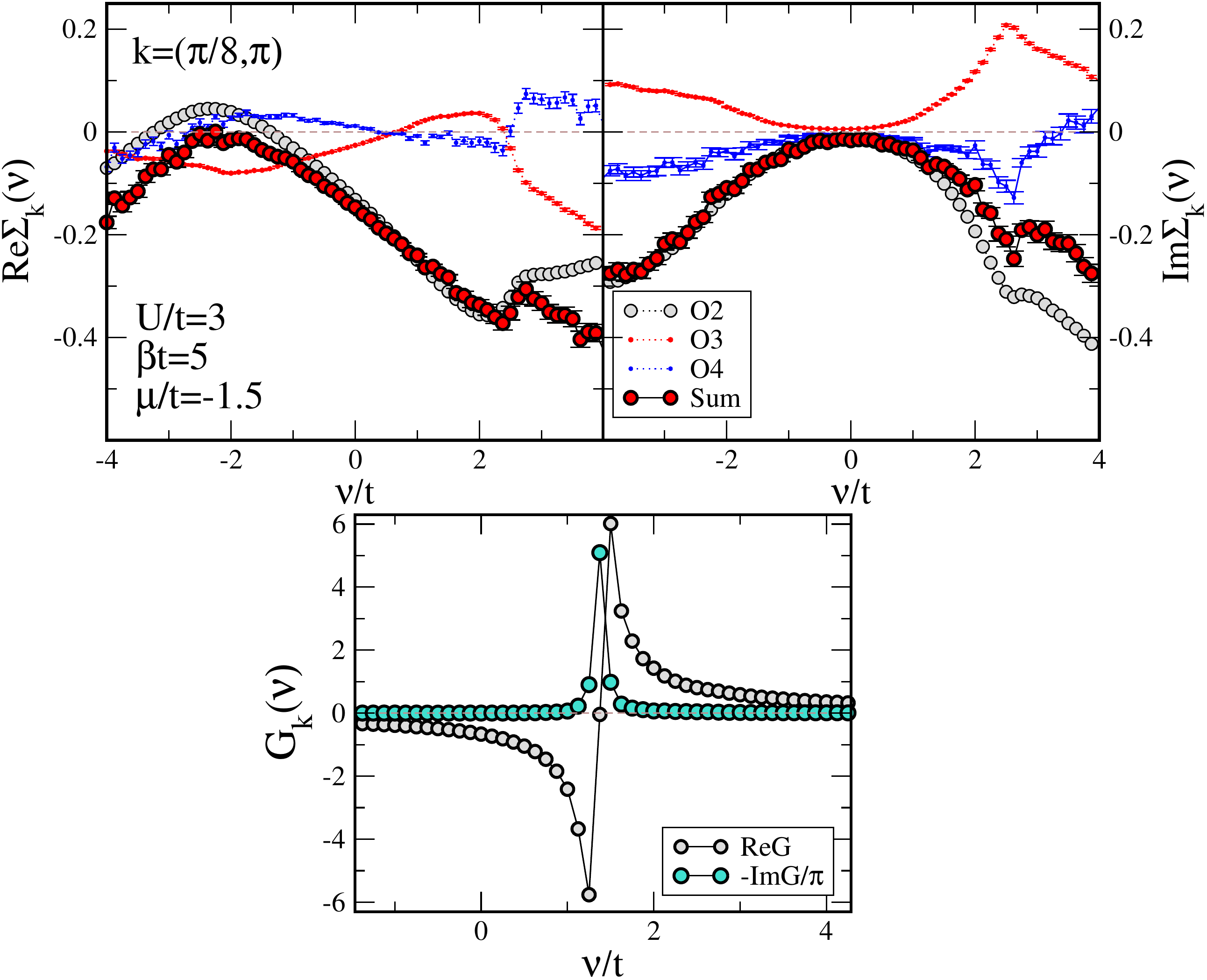}
        \caption{\emph{Top}: Real and imaginary parts of  self-energy, \emph{Bottom}: Green's function up to fourth order vs. real frequency $\nu$. Data are for parameters $U/t=3$, $\mu/t=-1.5$, $\beta t=5$ at $\textbf k=(\pi/8,\pi)$. We set $\Gamma/t=0.05$ in symbolic analytic continuation process $i\nu_n \to \nu + i\Gamma$. \label{fig:mf}}
\end{figure}

We show another example away from half-filling in Figure~\ref{fig:mf} but now evaluated on the real frequency axis at $U/t=3$ for $\beta t=5$ and $\mu/t=-1.5$.  
Since this parameter choice is within the convergence criteria mentioned in Figure~\ref{fig:doped}, we therefore expect small truncation errors and our results up to 4th order illustrate the utility and rather high accuracy attainable with the method.
For completeness we show the real and imaginary parts of the self-energy and each contribution from $m$th order ($Om$) as well as the sum up to 4th order. At each order the result is evaluated by forming subclasses and evaluating all diagrams of a subclass together. 
Of interest is the partial cancellation between 3rd and 4th order contributions for much of the frequency range. As a result, by comparing the $O2$ result to the entire sum we note a wide range of frequencies $\nu=-1\to2.5$ where both the real and imaginary parts of the 2nd order diagram are nearly equivalent to the sum up to 4th order.  Also shown in the lower panel are the real and imaginary parts of the Green's function resulting from the self-energy sum in the upper panels. One notes a typical form of the Green's function and can identify at which frequency ${\rm Re}G$ changes sign which corresponds to the energy $\epsilon(\textbf k)-\hbox{Re}\Sigma_{\textbf k}$. The form of the Green's function is essentially independent of $\Sigma$ far away from this boundary. Thus only self-energies near this boundary, here from $\nu\approx 0.5\to 2$, need to be evaluated to correctly represent the Green's function.

Essential in obtaining these results is managing the divergences of the AMI integrands that arise for evaluation on the real frequency axis.  While analytically these divergences always have cancelling terms each individual term might cause numerical overflow that must be managed. To do so, we use two regulators: an intrinsic scattering rate $\Gamma$ for the analytically continued frequency $i\nu_n\to\nu+i\Gamma$ that provides a width to the imaginary parts of the Green's functions; and a thermal regulator $\eta$ which enters the bosonic distribution functions in the $E\to0$ limit.  The constraint on these regulators for numerical correctness is that they be much smaller than the dominant energy scale, $\Gamma << \nu$ and $\eta << k_B T$. Operating outside this constraint will typically result in overly smoothed results, or reduced numerical values.  Our calculations are performed with $\Gamma/t=0.05$ and $\eta=10^{-5}$, though somewhat larger values can be used to improve statistical uncertainty without visible change to the result.
In addition to these regulators, for some diagrams there may exist terms in the AMI integrand that have no external frequency in the denominator and only a linear combination of energies.  This results in a large number of spurious poles inside the integration space of size $mD$ for $D$ dimensions that are not regulated by $\Gamma$ nor by $\eta$.  One needs only to avoid the direct evaluation of the integrand at these poles to obtain correct results. To do so it is \emph{essential} that the momentum integrals \emph{not} be performed on a regular $L\times L$ grid.  Doing so virtually guarantees evaluation of the integrand directly on a pole. Instead, sampling the space via MC methods by choosing random points in the integration space makes it unlikely to encounter these spurious poles. In addition, this extends the calculation to be effectively continuous in momentum space and provides results directly in the thermodynamic limit.

In a recent work, Vu\v ci\v cevi\'c and Ferrero\cite{ferrero:2019} have devised an alternate method of diagram evaluation starting from Eq.~(\ref{eqn:each_diag_in_goal}) but they first replace the product of bare Green's functions with a summation by employing a generalization of a partial fractions decomposition. In that work they allude to a number of obstacles that we do not seem to encounter. We suspect that the process of breaking the integrand into partial fractions produces many cancelling terms resulting in an unnecessary inclusion of many removable divergences. Avoiding this procedure as well as avoiding the use of a regular $L\times L$ grid, as we have done, has allowed us to use very small regulators such as $\eta/t=1e^{-4}\to 1e^{-8}$ in fermi/bose distribution functions and $\Gamma/t=0.05\to 1e^{-4}$ without particular difficulty.  This means we can correctly perform the $\Gamma \to 0$ limit required for true analytic continuation. 

\begin{figure}
\centering
    \includegraphics[width=0.95\linewidth]{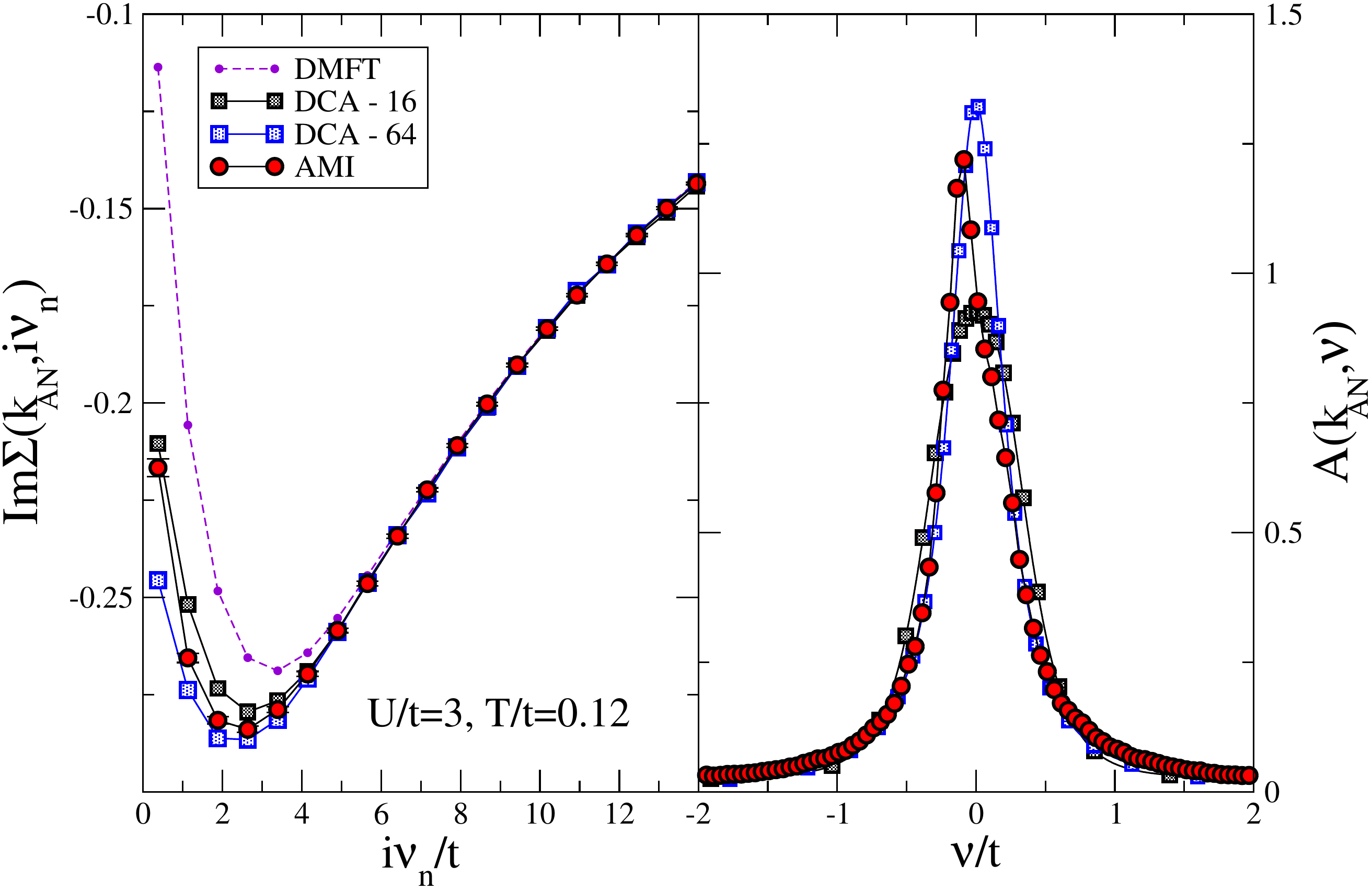}
        \caption{\emph{Left}: Imaginary part of the self-energy on the Matsubara axis at $\textbf k_{AN}=(\pi,0)$ for $\mu=0$, $U/t=3$, and $\beta t=8.33$.  Results from DMFT are shown as well as DCA data for 16 and 64-site clusters. \emph{Right}: Spectral function $A(\textbf k_{AN},\nu)$ on the real frequency axis. DCA results obtained via maximum entropy inversion.\cite{Levy2016}  AMI results assume $\Gamma/t=0.05$ in symbolic analytic continuation process $i\nu_n \to \nu + i\Gamma$. \label{fig:dcacompare} }
\end{figure}

Moving forward, we restrict our calculations to the half-filled model where we make use of the full power of the GIT methodology described in Sec.~\ref{sec:filters}. We present results truncated again at only 4th order. We need to evaluate only $N_r^{(m)}$ diagram groups from the last row of Table~\ref{tab:summary}. This amounts to only evaluating five diagrams, which can be accomplished extremely quickly. We also provide comparison to established numerical methods DCA and $\Sigma$DDMC.\cite{gukelberger:2017,Evgeny:2019}

We show in the left-hand frame of Figure~\ref{fig:dcacompare} imaginary part of the self-energy vs. $i\nu_n$ Matsubara frequencies obtained by the direct evaluation of the diagrams up to 4th order using AMI at the antinodal point $\textbf k_{AN}=(\pi,0)$ for $U/t=3$ and $\beta t=8.33$. 
The results are in perfect agreement with DCA after only a few frequencies,  $(i\nu_n>i\nu_5)$. This is expected since a larger value of $i\nu_n$ strongly suppresses high order contributions, reducing the truncation error sharply. Comparison to DCA at low frequency shows that the 4th order truncated series is surprisingly competitive with 16$\to$64-site DCA.\cite{yamada:4thorder,gebhard:4thorder} In general we expect our truncation error to grow for decreasing $i\nu_n$ or decreasing temperature, and here the error bars only reflect statistical uncertainty and do not represent truncation errors.

The power of AMI becomes apparent in the right-hand frame of Figure~\ref{fig:dcacompare} where we plot the real frequency spectral function at the antinodal point.  Recall that for AMI the analytic continuation involves only a symbolic replacement of $i\nu_n \to \nu +i\Gamma$ for some sufficiently small value of $\Gamma$. The resulting AMI spectral density is shown in red circles. For comparison we perform the numerical analytic continuation\cite{Levy2016} for the DCA Green's functions at the antinodal point. Surprisingly we see that the DCA result after numerical analytic continuation has the same broadening as determined by AMI directly on the real-frequency axis. One notes a slight asymmetry in the AMI result, evaluated at $\mu=0$.  This is due to a small value of ${\rm Re}\Sigma(\textbf k_F,0)\neq0$. This truncated expansion is not particle-conserving and therefore this represents a density very close to, but not equal to half-filling. Further, the AMI result near $\nu \to 0$ may be underestimated due to the energy and thermal regulators. One would need to maintain that $\Gamma <<\nu $ and further that $\Gamma << \hbox{Im}\Sigma(\nu \to 0)$ in order to guarantee correctness.  These considerations have not been addressed for this simplistic example.
\begin{figure}
\centering
\includegraphics[width=0.95\linewidth]{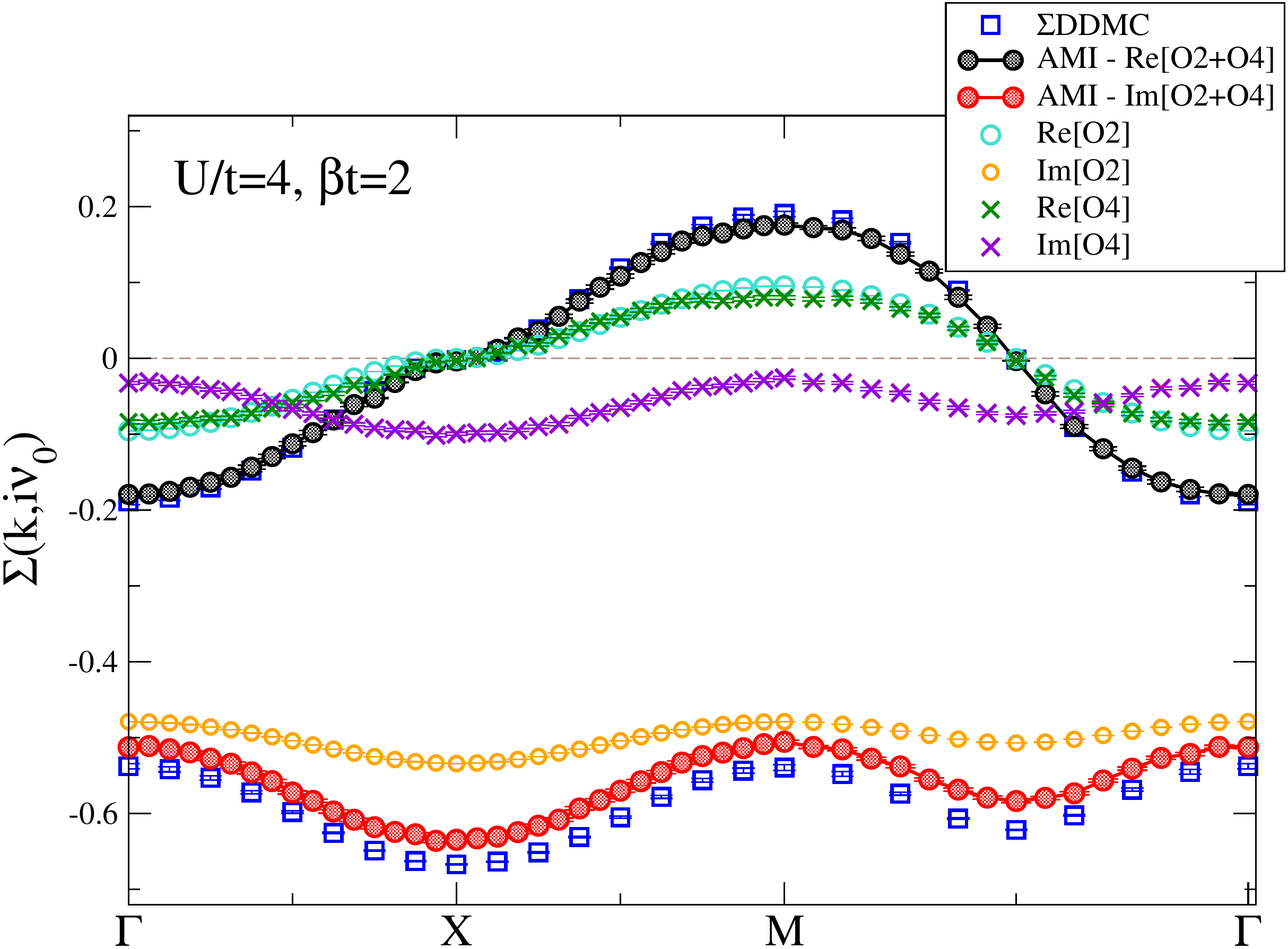}
        \caption{Real and imaginary parts of the self-energy at $i\nu_0$ through high-symmetry cuts in $k_x-k_y$ plane for $\mu=0$, $U/t=4$, and $\beta t=2$.  Upper/lower blue squares are the real/imaginary $\Sigma$DDMC results from Refs.~\onlinecite{gukelberger:2017,simkovic2017determinant}.\label{fig:sigmaddmc_compare} }
\end{figure}

As a further benchmark, we compare to high order $\Sigma$DDMC results at half-filling.\cite{gukelberger:2017}  In Figure~\ref{fig:sigmaddmc_compare} we show the real and imaginary parts of the self-energy at the first Matsubara frequency for high-symmetry cuts through the Brillouin zone. For discussion purposes we plot both the individual 2nd and 4th order AMI results as well as their sums. We see that the 2nd order contribution to the imaginary part of the self-energy (orange circles) is much larger than the 4th order contributions (purple crosses). This is not the case for the real part of the self-energy where the 2nd and 4th order contributions are nearly equal suggesting that the convergent behavior of the real and imaginary parts of the coefficients need not be the same. In both cases the sum of these results are surprisingly similar to the numerically exact $\Sigma$DDMC results and both the real and imaginary parts have the correct qualitative momentum dependence. Visually it appears that the real-part is a better approximation.  If we scrutinize the results at the $\Gamma$ point where the real part has values $(-0.179\pm0.002)$ and $(-0.188\pm0.004)$ for AMI and $\Sigma$DDMC respectively we find a $\approx 5\%$ discrepancy.  Repeating this for the imaginary part with values $(-0.511\pm0.002)$ and $(-0.537\pm0.004)$ we find again a $\approx 5\%$ discrepancy, therefore the visual distinction is only a matter of scale, and we see that the  relative truncation error is in practice much less than our numerical estimate. In each case, it must be true that the sum of terms of order $m\geq6$ results only in these small differences. The results are not generally expected to be this accurate for all parameter choices and indeed at lower temperatures we find that the deviation increases. This behavior has also been observed in order-by-order expansions from diagrammatic treatment of the dual Fermion method.\cite{gukelberger:2017, iskakov:2016}

\section{Conclusion}
We have presented a general framework to evaluate Feynman diagrammatic expansions that can be applied to virtually any expansion with any interaction. Specifically, our method is applicable to any diagrammatic expansion composed of the bare Green's functions with any frequency-independent two-body interaction. 

Important features of our method can be summarized as follows. The Matsubara sums are performed analytically using AMI. This allows for the symbolic analytic continuation $i\nu_n \to \nu+i0^+$ without any ill-defined numerical procedure. The full symbolic result of AMI in principle enables us to exactly (up to machine precision) evaluate Matsubara sums of each diagram in the expansion at any temperature even at $T=0$ limit, which is not accessible in DiagMC methods. We also determine the pole structure of the integrand of the diagrams, which enables us to divide the diagrams into groups which contain nearly cancelling pairs. We therefore sample pairs of nearly cancelling diagrams as a whole in Monte Carlo integration instead of sampling the diagrams one-by-one, which leads to a substantial suppression to the sign problem. Further, in the special case when there is particle-hole symmetry the cancellations are {\em exact}, while other diagrams within each group are exactly equal. Moreover, despite the factorially growing cost, the AMI result and diagram groups can be easily stored, i.e., one needs to solve the problem up to momenta integrations only once.  

As proof of concept we presented the application of this method to the self-energy expansion of the Hubbard model on a 2D square lattice with nearest neighbor tight-binding dispersion at and away from half-filling. The resulting diagram groups are provided in the Supplemental Material up to 6th order and these groups are also valid for 1D or 3D systems as well as valid for the imaginary time representation. 
As evidence of utility we provided as well numerical comparison of the low order expansion to other numerical methods and found excellent results when within the convergent range of the series.

While the procedure is a major advancement in evaluating diagrammatics on the real frequency axis it does not address the factorially growing diagram space, which remains time consuming to evaluate.  Further, it does not address the fundamental sign problem inherent in the analytic AMI results and in many cases the average sign remains small after AMI and is not generally improved by grouping diagrams.   
Finally, while AMI allows for the evaluation of any Feynman diagram at any temperature it seems that reduced temperature causes the expansion to favor higher order terms.

\section{Acknowledgments}
JPFL acknowledges the support of the Natural Sciences and Engineering Research Council of Canada (NSERC), RGPIN-2017-04253. 
Computational resources were provided by ACENET and Compute Canada.  Our Monte Carlo codes make use of the open source ALPSCore framework\cite{ALPSCore,alpscore_v2} and we have used the open source code Maxent\cite{Levy2016} for numerical analytic continuation.

\bibliographystyle{apsrev4-1}
\bibliography{refs.bib}

\end{document}